\documentclass{article}
\usepackage{ijcai17}
\usepackage[linesnumbered,ruled,vlined]{algorithm2e}
\usepackage{booktabs}
\usepackage{tabularx}
\usepackage{graphicx}
\usepackage{url}
\usepackage[utf8]{inputenc}
\usepackage{breqn}
\usepackage[T1]{fontenc}
\usepackage{cuted}

\newcolumntype{C}[1]{>{\centering\let\newline\\\arraybackslash\hspace{0pt}}m{#1}}
\newcommand{\eg}{{\em e.g.}}
\newcommand{\ie}{{\em i.e.}}
\newcommand{\etc}{{\em etc.}}
\usepackage{amsopn}

\newcommand{\argmaxE}{\mathop{\mathrm{argmax}}}

\def\etal{\textit{et~al.}\xspace}

\begin{document}

\title{Creative Exploration Using Topic Based Bisociative Networks}

\author{Faez Ahmed\\ 
Dept. of Mechanical Engg.\\
University of Maryland\\
faez00@umd.edu
\And
Mark Fuge\\ 
Dept. of Mechanical Engg.\\
University of Maryland\\
fuge@umd.edu
}

\maketitle

\begin{abstract}
 
Bisociative knowledge discovery is an approach that combines elements from two or more ``incompatible'' domains to generate creative solutions and insight.
Inspired by Koestler's notion of bisociation, in this paper we propose a computational framework for the discovery of new connections between domains to promote creative discovery and inspiration in design. Specifically, we propose using topic models on a large collection of unstructured text ideas from multiple domains to discover creative sources of inspiration.
We use these topics to generate a Bisociative Information Network\textemdash a graph that captures conceptual similarity between ideas\textemdash that helps designers find creative links within that network. 
Using a dataset of thousands of ideas from OpenIDEO, an online collaborative community, 
our results show usefulness of representing conceptual bridges through collections of words (topics) in finding cross-domain inspiration.
We show that the discovered links between domains,
whether presented on their own or via ideas they inspired, are perceived to be more novel and can also be used as creative stimuli for new idea generation.
\end{abstract}

\section{Introduction}

In the field of design and engineering, many existing tools support creativity during idea generation. These tools help designers generate solutions and explore a larger design space. This exploration is useful during conceptual design when divergent thinking can help avoid fixation \cite{jansson1991design,purcell1996design} and dead-end branching~\cite{Shah2003111}, leading to more creative solutions~\cite{siangliulue2015toward}. 
But how should one go about this creative exploration? One option is to combine ideas from different sources.

Henri Poincar{\'e} \cite{poincare1910mathematical} said ``Among chosen combinations the most fertile will often be those formed of elements drawn from domains which are far apart\ldots Most combinations so formed would be entirely sterile; but certain among them, very rare, are the most fruitful of all.''
Inspired by similar thinking, designers often connect seemingly unrelated information, for example, by using metaphors or analogy~\cite{hey2008analogies}. The entire field of Biomimicry or Biologically-inspired Design\textemdash finding designs in nature and adapting their principles to man-made artifacts\textemdash is just one link between different domains that designers have found fruitful. This paper presents computational techniques for finding other such bridges between domains\textemdash or \textit{bisociations}~\cite{koestler1964act}\textemdash given a set of design ideas.

Identifying good bisociations requires answering three technical questions: 1)~What (specifically and computationally) does it mean to ``bridge'' a domain? 2)~Assuming that I find such bridges, which bridges are creative? and 3)~Assuming I have found a creative bridge between domains, how do I represent that conceptual bridge to a designer such that they find it useful?

Researchers have heavily studied those first two questions. The remainder of the introduction reviews that past work and then focuses on the main theory that we leverage in this paper\textemdash Bisociative Networks\textemdash which builds a network (\ie, graph) and then uses properties of that network to find bridges between domains. However, the way standard Bisociative Networks represent ideas (\ie, using \textit{bridging words}) causes problems for design exploration, both when forming the bisociative network and when using the output.
This paper's below methodology resolves those two problems. We use a randomized controlled experiment and qualitative comparisons of the output to demonstrate its efficacy with respect to standard baselines and existing state of the art.

\subsection{Creativity and Finding Inspiration}
\cite{boden1994precis} defines creativity as ``the ability to come up with ideas or artifacts that are new, surprising and valuable.'' 
The first factor\textemdash newness or novelty\textemdash implies that an idea should not have existed previously, \ie, be original.

The second factor is the notion
of surprise\textemdash an idea may be surprising because it may seem unlikely or unfamiliar (even if it is not, in and of itself, new).
The third factor is the notion of value\textemdash a new concept must be valuable (not just new and surprising) to qualify as creative.
However, an idea's value depends on many scientific, social, economic, political and other factors. Hence agreeing over the degree of creativity is difficult and context dependent.

Boden's model of creativity also defines three roads to creativity: 
Combinatorial Creativity which combines ideas within a domain, Exploratory Creativity which finds new ideas across existing domains, and Transformational Creativity which finds something new outside known domains. We focus on Exploratory Creativity\textemdash coming up with a new meaningful combinations. An example of such creative inspiration is 
the design of the Shinkansen high speed train in Japan, which was inspired by the beak of a kingfisher \cite{deyoung2009discovery}. 
Similarly, natural silk inspired the design of synthetic fibers, such as Nylon and Kevlar \cite{gosline1999mechanical}.

To support Exploratory Creativity, researchers across many fields have developed different computational approaches under different names. 
In the engineering design domain, predictive
models have been employed  to characterize
hidden patterns within existing data sets. 
For example, \cite{benami2002creative} investigated factors which stimulate creativity in conceptual design. \cite{pahl2007generic} presented a generic model of the process leading to innovative design by comparing all the processes of creating outlined in the psychological
literature. Their
model defines and makes visible the path of generation and divergence of ideas, followed
by a period of `editing' and a final convergence into innovation. 
\cite{zahner2010fix} provide two methods to reduce fixated thinking--- abstracting and re-representing. They showed that abstractness
promoted original ideas in the design of information systems. Similarly, the effect of different level of abstraction for textual representations in \cite{gonalves2012find} showed the benefit of distant textual stimuli
for generating original ideas.
These models, while useful for single domain, are often limited in their ability to draw connections between seemingly unrelated domains. Sometimes, innovative design solutions across multiple, seemingly unrelated domains may be omitted entirely.

Generating original solutions by borrowing ideas from multiple domains has been a key challenge for designers. One of the widely used method to this problem is Design-by-analogy (DbA), which has been shown to be an effective method for inspiring innovative design solutions. It is a practice in which designers use solutions from other domains to gain inspiration. DbA supports designers in developing conceptual designs for new products by discovering new insights from multiple domains. Engineering designers have often used design by analogy for Bio-Inspired design too. It allows engineers to take ideas from nature
and develop new design solutions for engineering problems by searching design analogies from biological domains. However, identifying useful solutions from outside domains using DbA is non trivial. For example,
\cite{fu2013meaning} measured distances between patents using a hierarchical Bayesian model and showed that priming people with patents too ``far'' (in terms of tree path length) from a target patent can be harmful to retrieving analogies, while likewise recommending patents too ``near'' can result in design fixation. Likewise, \cite{chan2014best} analyzed winning ideas submitted to the online design challenge website OpenIDEO and found, via each idea's citations network, that the best design ideas often came from sources of inspiration which are not far away (in terms of path lengths in the citation graph). 

Like these prior approaches, our method provides an automated computational tool to find abstract inspirations from unrelated domains by modeling a concept of ``distance" between ideas and domains. However, unlike past approaches, we do not assume that this distance metric is context independent (unlike, for example, tree hierarchy or citation graph paths). 
Our method differs from past studies, as it does not directly measure distances to identify ideas that are far-off or close by to a domain,
but learns from the data to identify possible sources of inspiration.
It does so by looking at ideas which are confused to belong to some other domain. Whereas past approaches find existing ideas as creative inspiration, we discover hidden concepts within ideas, which act as creative inspiration.
Specifically, we focus on a prior line of work called Bisociative Creative Information Exploration which is inspired by Koestler's model of creativity~\cite{koestler1964act} proposed in the 1960s. His model centers around the concept of \textit{bisociation}. 

\subsection{Koestler's Concept of Bisociation}

Bisociation, according to Koestler, means joining unrelated, often conflicting, information in a new way \cite{koestler1964act}.
He makes a clear distinction between habitual thinking (association) operating within a single plane of thought, and the more creative bisociative mode of thinking which connects independent planes of thought. Koestler conjectured that bisociation is a general mechanism for the creative act in the field of humor, science, engineering, and the arts. 

\begin{figure}
\centering
\includegraphics[width=0.9\columnwidth]{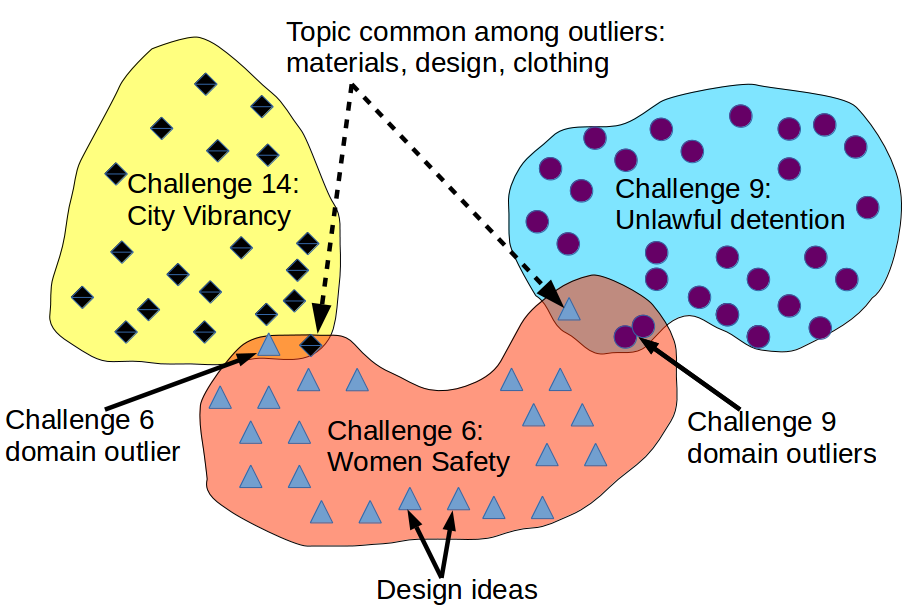}
\caption{Three design domains and outlier ideas (ideas from these domains which are more similar to other domains). Topics common among outlier ideas but uncommon overall have high bisociation score. In this example, topic on using `clothing' and `material' is a b-topic. These ideas and challenge domains were sampled from the OpenIDEO dataset we introduce in the Results section.}
\label{fig:visual}
\end{figure}

Figure~\ref{fig:visual} provides a simple illustration to bisociation, by showing examples from a set of three OpenIDEO design challenges (or what we will call ``domains''). Specifically, each of the three challenges\textemdash improving womens' safety in urban environments, reducing the risk of unlawful detection, and increasing vibrancy in cites facing economic decline\textemdash had several ideas (represented as markers in Fig.~\ref{fig:visual})  that users submitted in response to the challenge. \textit{Bisociations} are indirect connections between ideas, which cross the border between (\ie, bridge) domains (or in this case, different challenges). On surface, these domains are quite dissimilar, however, one may discover an indirect concept, which is common to these domains. In Fig.~\ref{fig:visual}, for example, several ideas across all three challenges leverage the idea using wearable accessories to address the challenge problem. Koestler would call that concept\textemdash the use of wearable accessories\textemdash a bisociation, in that ideas from one challenge or domain might more easily transfer between domains that jointly share the concept. We show later in the paper how ideas from a design collection for these three domains were found to have this concept common among them. We will also define precisely how to represent and compute bridging concepts in later sections.

Researchers have applied bisociation most readily to the discovery and exploration of research literature (\ie, academic papers). This work is typified by the work of \cite{swanson1986fish} who introduces ``Swanson linking'' to connect medical literature by assuming that new knowledge and insight arises when connecting knowledge sources which were thought to be unrelated. 
In his seminal paper, Swanson investigated connections between migraine and magnesium, based on published research papers. He found indirect relations via \textit{bridging terms (b-terms)}\textemdash words that signaled possible connections between two domains\textemdash that suggested magnesium deficiency may cause migraines. 

Several researchers have continued Swanson's line of research, for example the RaJoLink method \cite{petriue2009literature} and the BISON project \cite{berthold2012towards,dubitzky2012towards}. These efforts pursued new algorithms to create, analyze, and explore domain-bridging words within text document collections. For example, \cite{jurvsivc2012bridging} select and rank keywords they claim highlight bridging words that help people discover cross-domain links that can lead to new ideas. They show that their methodology places a significantly higher number of bridging words towards the top of a ranked list compared to chance rankings.  \cite{kang2016bisociative} and \cite{kang2017exploring} applied bisociative design methodology to discover product attributes that correlate to an increase in enterprise profit. They do so by analyzing the 
associations between function attributes and 3D form attributes among different products. They use Latent Dirichlet Allocation (LDA) to extract the function attributes from the product descriptions and Reeb graphs to represent the form. 
\cite{tucker2012bisociative} studied bisociations by decomposition of a design artifact by form, function and behavior to quantify the level of similarity among items across domains. The authors demonstrated their method on marine and aviation domains. These bisociation studies are also supplemented by research on bisociative networks, predominantly used for creative exploration.

\subsection{BisoNets: Bisociative Information Networks}

A Bisociative Information Network (or BisoNet) is a method of practically computing Koestler's bisociation. A BisoNet represents concepts as a network\textemdash a graph with vertices and edges\textemdash and then defines functions on that network that compute creative bisociations. This approach leads to two intimately coupled technical challenges. The first, which is common to all network modeling approaches, lies in how one constructs the network itself\textemdash \eg, what are the nodes and edges in the network, and how does that choice affect the outcome? The second, specific to bisociation, lies in how one computes which nodes ``bridge'' domains in a creative way. We review the first challenge in this section, and address the second later in our methodology. In brief, the way that past work represented BisoNets (\ie, using specific bridging words), while effective for academic literature, does not perform well for design concepts. One of the contributions of this paper lies in defining a more appropriate network representation and subsequent function for computing bisociation under that new representation.

Specifically, a BisoNet is a weighted, undirected, k-partite graph\footnote{A $k$-partite graph is one whose nodes can be one of $k$ number of types, and whose edges only link nodes of different type\textemdash \eg, assigning papers to reviewers is a bipartite graph, where a paper can link to a reviewer, and a reviewer to a paper, but reviewers cannot link to other reviewers.} of concepts, such that similar concepts are connected by an edge\textemdash in essence, a similarity graph, but with a particular form of similarity called \textit{bisociation} that we detail later in the paper. 
Vertices in BisoNets can represent any unit of analysis, such as words, documents, ideas, people, \etc~Vertices of the same type are grouped into vertex partitions\textemdash for example, partitioning all words from a particular document together, or partitioning all articles from a given field together. 

As with all network models, a key differentiator among past work lies in how they calculate the edge weight between the graph nodes. For example, finding relations between such nodes often focuses on discovering semantically related terms, frequently using lexical databases and ontologies. Edge weights can be calculated using measures like cosine similarity, Normalized Google Distance Measure (NGD) \cite{cilibrasi2007google}, or similarity functions tailor-made for bisociation discovery, like  Segond and Borgelt's Bison measure \cite{segond2009bisonet}.

Researchers have applied BisoNets to exploration of Biological and Financial Literature \cite{schmidt2012bisociative} and Music Discovery \cite{stober2012bisociative}, with unstructured text documents being one of the most widely used (and most challenging) applications. These past text-based approaches work well when there are specific, technical terms embedded in documents that are shared between domains. For example, in the autism-calcineurin benchmark dataset~\cite{berthold2012towards}, standard BisoNet exploration tools identify individual scientific terms like ``paroxysmal'' or ``Bcl-2'' that discover links between two scientific domains (in that case, between autism and the human immune system). 
However, this example also highlights two key issues with past BisoNet approaches that make them ill-suited for creative design exploration.

First, past representations relied on identifying specific bridging \textit{words}. As we demonstrate in our results, for design concepts this does not work well since design descriptions often rely on multiple words or ideas taken holistically together as a system\textemdash that is, there are no magic bridging words (like ``Bcl-2'' in the autism example) but rather collections of words or phrases that, in aggregate, provide a new frame within which to view a design problem. Current approaches to bisociation do not handle such cases. This paper resolves that problem by introducing bridging topics---called b-topics, rather than bridging terms, to capture richer representations for bridges across domains.

Second, existing BisoNet approaches find bridging terms between only two, pre-identified domains (\eg, the autism-calcineurin or migraine-magnesium datasets). This assumes that one knows, apriori, which two domains will likely produce good bridging terms. While this pre-knowledge of domains may exist for certain design applications (\eg, in bio-inspired design), in general we largely do not know which combinations of two domains will be fruitful. This paper resolves this problem by generalizing existing bisociation techniques to exploration across multiple domains at once, not just between two. We demonstrate below that this leads to much richer exploration of possible bridging concepts than if we were to pre-select two domains apriori.

Aside from bisociation, some researchers have approached the same problem from the perspective of Serendipity \cite{roberts1989serendipity,kamahara2005community}.  Serendipitous discoveries overlap with bisociations since they often involve realizing a connection between dissimilar domains of knowledge. Serendipity has mainly been applied to recommender systems \cite{onuma2009tangent}. 

\subsection{This Paper's Contributions}

Our work builds upon earlier BisoNet techniques~\cite{schmidt2012bisociative}, but with three main differences.
First, we apply bisociation principles to fourteen domains that are broader than analysis of scientific papers. Most of the previous techniques 
applied BisoNets to either a migraine and magnesium dataset or an autism-calcineurin domain. These datasets only have two domains and performance evaluation is straightforward due to the advantage of having gold standard bridging terms. Second, by comparing with CrossBee tool\textemdash the existing state of the art in computational bisociation\textemdash we show that finding words as b-terms for unstructured text is not as useful for design concepts, thus small collections of words should be used instead. We propose using topic models for this purpose and re-define bisociation metrics such that they work for topics. Finally, we evaluate our method using human preferences elicited by crowd workers on Amazon Turk. 

This paper's key contributions are:
\begin{enumerate}
\item The introduction of bridging topics\textemdash via Topic Models~\cite{blei2009topic}\textemdash as a representation for computing bisociative links in the network.
\item Introducing a new objective function for 
ranking topics by their bisociation potential 
\item Generating a BisoNet from topic representations via identifying likely edges.
\item Demonstrating that bisociation can be used in domains far broader than identifying bridging words within academic literature. Such bisociation produces new inspirational frames for design problems that, within our experiments, led humans to generate more creative solutions.
\item Generalizing the principles of bisociation to simultaneously handle multiple domains, rather than just between two domains.

\end{enumerate}

One major challenge compared to past BisoNet work is the lack of comprehensive benchmark datasets for multiple domains outside of scientific literature. One of this paper's ancillary outcomes is to enable creation of such a dataset, so that others can study multi-domain bisociation in broad design domains.
We have made data corresponding to our results available online \footnote{\url{https://github.com/IDEALLab/bisonet}}.

\section{Methodology}

Let us say that we are given ideas from 
$D$ domains. Here, we propose a method, which finds a ranked list of concepts which indirectly connects these domains. Next, we also show how these concepts can be used to generate a BisoNet. Our approach to creative exploration consists of three main steps: 1) constructing the network nodes\textemdash that is, learning each idea's topic representations, 2) finding  likely \textit{bridging topics} (or \textit{b-topics}) and using those topics to create network edges that connect idea domains, and 3) constructing a BisoNet from highly probable bridging topics, so that users can explore and navigate a graph of cross-domain inspirations.

\subsection{Representing Ideas} 

The first step is to computationally represent an idea or design. In this research, we only consider text documents; however, the techniques we develop below for topic collections will transfer to other inputs or media (\eg, images). Traditionally, research on representing text documents largely used a vector space model where a document is expressed by a vector of keyword weights using the TF-IDF method \cite{salton1988term}. Researchers have since developed various other dimension reduction techniques to capture the hidden semantic structure in a document including probabilistic Latent Semantic Analysis (pLSA) \cite{hofmann1999probabilistic} and topic modeling \cite{blei2009topic}. 
The ``topics'' produced by topic modeling techniques are clusters of similar words. 
A topic model captures this intuition in a mathematical framework, which allows examining a set of documents and discovering, based on the statistics of the words in each, what the topics might be and what each document's balance of topics is. 

Latent Dirichlet Allocation (LDA)~\cite{blei2003latent}\textemdash is a widely used topic modeling method. In LDA, each document is described as a random mixture over a set of hidden topics and each topic is a discrete distribution over a text vocabulary\textemdash that is, words can belong to discrete clusters, and LDA learns from data how strongly any word should belong to any cluster. LDA has been applied to many areas \cite{wei2006lda,krestel2009latent,rosen2004author} and several variants of LDA have also been proposed to tackle correlated~\cite{blei2007correlated} and network-based~\cite{chang2009relational} structure between topics. In design, \cite{chan2014overreliance} used LDA to represent ideas on OpenIDEO and showed that concepts that cite sources had greater success than those that did not cite sources of inspiration. \cite{chan2015importance} also used LDA to represent ideas, where they hypothesized that iteration is necessary to convert far combinations into creative concepts.

The key insight from topic models relevant to our work is that, rather than using the specific bridging words from a document (as in standard BisoNet examples like ``Bcl-2'' above), we can instead cluster words together into overall topics that contain sets of related words. 
For example, single bridging word like `care' can be vague. It can refer to care in hospitals, care for the elderly or health insurance care etc. 
However, the ambiguity is reduced for a semantically related collection of words like \{`care', `health', `patient', `hospital', `doctor', `medical', `center'\}, 
as it provides clearer framing and context.
Although, subjectivity of interpretation is a desirable property of our approach compared to showing existing ideas, it can often act as double edged sword in the design process.
While showing existing ideas can often be too specific, showing single words can be too ambiguous. 
Under what conditions does including multiple words increase or decrease the clarity?
One way to think about context clarity is whether a word (or set of words) collapses the Conditional Shannon Entropy of the topic posterior probability distribution in a topic model. When single words are used, the topic distribution generally has high entropy, implying that single word can come from many topics or contexts. When multiple related words from the same topic are used, the posterior probability distribution collapses to zero entropy (there is no topic uncertainty) and thus refers to only one topic. The above assumes that topics are a reasonable proxy for ``context" or ``framing"\textemdash an assumption we believe is reasonable, given that topic models are designed to capture document context. 
Hence, we claim that collections of relevant words (\ie, topics) can act as better bridges between design domains than individual terms used in current Bisociative Networks.

In this paper, we use LDA to capture the topic distribution of ideas, however our contributions are independent of the specific topic model variant or implementation used. Specifically, we learn the topic distribution for each idea a corpus of designs\textemdash this means that we represent each design idea (text document, in this case) as a $M$-dimensional vector of numbers between $0$ and $1$ that corresponds to which topics are most prevalent in that idea. We use these vectors to identify edges and possible bridging topics.

\subsection{Bridging Topic Identification}
Given sets of ideas and their topics proportions, our goal is to find, for a given domain or set of ideas, what topics might bridge across other domains. One na{\"i}ve approach to finding bridging topics might be to simply look for topics that two or more domains have in common\textemdash after all, if a topic is highly represented within two domains, it seems sensible to expect that those topics would somehow bridge those two domains. The main problem with that approach is that the topics that are \textit{both} representative of a particular domain \textit{and} common across domains tend to be overly general topics that do not provide much creative insight\textemdash for example, common cross-domain topics might include topics such as \{`the', `and', `is', `of'\} or \{`man', `woman', `he', `she', `they'\}, \etc~While such topics certainly do bridge across domains, they are unlikely to meaningfully re-frame the problem in a creative way. 

Instead, we are looking for a kind of ``Goldilocks'' topic; topics that are uncommon enough to bring new insight to a problem, but common enough across domain outliers that the topic can be readily understood and adapted. 
This intuition\textemdash that we need to identify outliers within domains, but commonalities between domains\textemdash was the primary goal of previous research on bisociation; the central idea being to rank all bridging topics as a function of how rare they occur overall and how common they are among outliers within its own domain.

Specifically, we generalize the approach of Jur\u{s}i\u{c} \etal \cite{jurvsivc2012bisociative} to collections of word (topics) rather than single words. Their essential approach was to train a machine classifier to distinguish domains from one another using individual words within documents, and then search for documents or terms that the classifier reliably mis-classifies as a different domain. Why is that approach reasonable? The intuition is that documents that actually live within one domain\textemdash but are consistently classified as being in another (false negatives)\textemdash are more likely to ``bridge'' domains. Jur\u{s}i\u{c} \etal found this outlier-finding approach to be stable, even under minor modifications to the dataset, and that it consistently located the gold-standard bridging terms within the benchmark dataset.

To find bisociation scores for topics, we 
first find outlier ideas in every domain.
Here, outlier ideas are false negatives in the multi-label classification--- documents that have greater similarity to documents in some other domain than to those of their own domain. To find these outliers, we train a multi-class classifier and the documents wrongly classified by it (false negative) are marked as domain outliers.
The input to the multi-class classifier is the vectorial representation for each document and the output labels are the domain index. 
Ground truth during training is the true label of the domain. 
If one uses a poor classifier with large number of false negatives (low recall), it would wrongly consider many ideas as outliers.
Hence, the b-topic scoring will be erroneous and topics in such domains may get artificially high b-scores.

The rationale is that topics with high bisociation score are more common in outlier documents and less common overall. The outlier documents according to classification models should not belong to their domain and thus are likely to have borrowed concepts from other domains. Let ${\cal I}$ be the set of all $N$ documents from $M$ domains and $O_d$ be the set of outliers for domain $d$. Let $X$ be the $N\times T$ document-topic matrix, where row $i$ represents $i^{th}$ document's $T$ dimensional topic proportion vector. For topic $t$ in domain $d$:
\begin{equation}
\mbox{Topic bisociation score (t,d)}=\frac{\sum_{j\in O_d} X_{j,t} }{\sum_{i\in {\cal I}} X_{i,t}} 
\label{eq:btopic}
\end{equation}

The above score is used to rank every topic by their potential to be a true  bisociation candidate for a given domain ($d$). 
For classification with multiple domains, we build a joint classification model to simultaneously classify all the documents. To make the outlier set more robust, the prediction scores for the three classifiers with highest accuracy are added to find the output domain in classification. Documents whose predicted challenge are different from true challenge are allotted to the outlier set. While we describe the exact classifiers we use in our results section, the specific choice of classifier is not central to the contributions of this paper; any ensemble that meaningfully reduces the classifier variance should suffice.

\begin{algorithm}
\footnotesize
\caption{{\sc RankBisociativeTopics}}
\DontPrintSemicolon 
\KwIn{\\A collection of domains ${\cal D}$ (with $|{\cal D}|=N$)\\
		A collection of ideas ${\cal I}$\\
        vector $\mathbf{d_{\cal I}}$ of which domain $d\in {\cal D}$ each idea $i\in {\cal I}$ belongs to\\
        A query domain $q \in {\cal D}$}
\KwOut{List of topics ranked by Bisociation w.r.t. domain $q$} 
\SetKwFunction{lda}{runLDA}
\SetKwFunction{classifier}{trainDomainClassifier}
\SetKwFunction{predict}{predictDomainProbabilities}
topics, $X \gets${\sc VectorizeIdeas(${\cal I}$)}\;
$O \gets${\sc FindOutliers($X,\mathbf{d}_{\cal I}$)}\;
topicScores = $\sum_{j\in O_d} X_{j,t} / \sum_{i\in {\cal I}} X_{i,t}~~\forall t\in T$\;
\Return topics.rankBy(topicScores)\;
\;
\SetKwProg{Pn}{Function}{:}{\KwRet}
\SetKwProg{Def}{def}{:}{}

\Def{\sc VectorizeIdeas(${\cal I}$)}{
topics, X = \lda(${\cal I}$)\;
\Return topics, X\;
}

\Def{\sc FindOutliers($X,\mathbf{d}_{\cal I}$)}{
$classifier = $\classifier($X$,$\mathbf{d}_{\cal I}$)\; 
$\mathbf{D}_{predicted} = classifier.$\predict($X$)\;
$outliers \gets \emptyset$\;
\For{$i \in X$}{
	$d_{true} \gets \mathbf{d_{\cal I}}[i]$\;
    $d_{predicted} \gets \argmaxE_{d\in {\cal D}}  \mathbf{D}_{predicted}[i,d]$\;
    \If{$d_{true} \neq d_{predicted}$}{
    	$outliers \gets outliers\cup i$
    }
}
\Return outliers\;
}
\label{algo:topic}
\end{algorithm}

\begin{algorithm}
\footnotesize
\DontPrintSemicolon 
\KwIn{\\A collection of domains ${\cal D}$ (with $|{\cal D}|=N$)\\
		A collection of ideas ${\cal I}$\\
        vector $\mathbf{d_{\cal I}}$ of which domain $d\in {\cal D}$ each idea $i\in {\cal I}$ belongs to\\
        A query domain $q \in {\cal D}$\\
        (Optional) Bisociation threshold $\tau$ for a topic to become a vertex\\
        (Optional) Bison Measure threshold $\epsilon$ for two topics to share an edge
        }
\KwOut{BisoNet ($G$)} 
\SetKwFunction{lda}{runLDA}
\SetKwFunction{classifier}{trainDomainClassifier}
\SetKwFunction{predict}{predictDomainProbabilities}
\SetKw{Continue}{continue}
$V\gets \emptyset,E\gets \emptyset$\; 
\For{$q \in {\cal D}$}{
	$T_q$ = {\sc RankBisociativeTopics}$({\cal D},{\cal I},\mathbf{d}_{\cal I},q)$\;
    $T_q$ = {\sc DropLowBisociatonTopics}($T_q,\tau$)~~~(Optional, for pruning)\;
    $V\gets V \cup T_q$
}
\For{$T_i, T_j \in V$}{
	$t = {\sc topicBisonMeasure}(T_i,T_j)$~~(Eqn.~2)\;
    \If{$t < \epsilon$}{
    	\Continue~~(Optional, for pruning)
    }
    $E \gets E\cup (T_i,T_2,j)$
}

\Return{$G=(V,E)$}\;
\caption{{\sc GenerateBisoNet}}
\label{algo:change}
\end{algorithm}

\normalsize
\subsection{Generating the BisoNet}

Lastly, we create a BisoNet where links between bridging nodes in different domains can be visualized and understood using graph exploration techniques. Essentially, we define a procedure for linking the bridging topics (b-topics) of a BisoNet by finding weights that indicate the association strength. 

For BisoNets with words as nodes (rather than topics), \cite{segond2009bisonet} showed that keeping the edges between words that had the highest bisociation scores performed well at bisociative discovery\textemdash they referred to the ranking procedure as a \textit{Bison Measure}. 
We modify their proposed Bison measure to use topic proportions instead of term frequencies, applying the same rationale for topics and define the Topic Bison Measure $T(p,q| D_1,D_2)$ between two topics p and q as:
\begin{strip}
\begin{equation}
{\mbox{T}_{D_1,D_2} (p,q)=\sum_{i\in R}{\Bigg( \underbrace{\sqrt[k]{ X_{i,p}}
\cdot \\
X_{i,q}}}_{\mathrm{Both~topics~occur~in~idea?}}\cdot  \underbrace{\bigg(1-\frac{\mid \tan ^{- 1} (X_{i,p})-\tan ^{- 1} (X_{i,q}) \mid}{\tan ^{ - 1} (1)}\bigg)}_{\mathrm{Topic~proportions~similar?}}\Bigg)} 
\label{eq:2}
\end{equation}
\end{strip}

Where $R$ is the set of $i$ ideas obtained by the union of domains $D_1$ and $D_2$. In Equation~\ref{eq:2}, the product term $X_{i,p}\cdot X_{i,q}$ implies that for two topics to be similar (have a high bison measure), they should both have large positive proportions in a document, as a vanishing topic proportion means that the two topics do not co-occur in the corresponding document.
Secondly, they are more related if they have similar proportions. To further understand this, we have to keep in mind that having two topics, both of which have a topic proportion of 0.1, should be less important than having two topics with a topic proportion of 0.5. In the first case, the topics we are comparing appear only rarely in the considered document. On the other hand, in the latter case these topic appear very frequently in this document, which means that they are strongly linked according to this document.

The arctan function normalizes the effects of comparing topic proportions of different magnitude. Parameter $k$ can be adjusted according to the importance one is willing to give to low topic proportion values. 
Hence, this form has the advantage that it takes into account that two topic proportion values for the same index have to be positive, similarity between topics is greater if the topic proportion values are large and the same difference between topic proportions has different impact according to the values of the topic proportions.

The procedure to obtain a BisoNet is described in Algorithm~\ref{algo:change}. 
To simplify the network for visualization purposes, one may threshold the bisociation score of topics to select a small percentage of the highest scoring topics as edges;
these topics have high potential to be bisociative. After calculating the edge weights (the topic bison measure) between remaining topics, edge pruning
can be done to retain only a small fraction of highest weight edges.

\section{Results and Discussion}

\begin{figure}
\includegraphics[width=\columnwidth]{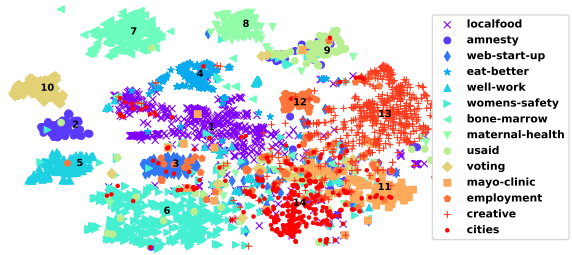}
\caption{All ideas from 14 challenges projected on a 2-D plane using t-Distributed Stochastic Neighbor Embedding (t-SNE). Some challenges (\eg, the voting challenge \#10), do not overlap many domains, while others (\eg, \#14) may have significant overlap.}
\label{fig:tsne}
\end{figure}

To study our method's effectiveness on a concrete example, we apply our technique to 14 OpenIDEO challenges to find interesting connections between domains. We then create a BisoNet for graph exploration and show meaningful themes discovered between different domains. Finally, we verify our results with different human experiments conducted with crowd workers.

\subsection{Dataset}
OpenIDEO is a successful online open innovation community centered around designing products, services, and experiences that promote social impact by building of ideas from distributed individuals \cite{fuge:openideo_JCISE_2014}. Generally, challenges have various stages like: `Research, Ideas, Applause, Refinement, Evaluation, and Winners.' and address very different social issues.
We focus on the `Ideas' stage, where participants generate and view potential design ideas. In this stage, hundreds to thousands of ideas are submitted in a single challenge. Reading ideas posted in past challenges or even the same challenge to gain inspiration when developing their own ideas is challenging\textemdash for a single, medium-sized challenge ($\approx 500$ ideas) it would take a person over 40 hours to read all idea entries. Because of this, participants often filter by date, the total number of comments, or just pick ideas randomly from the same challenge as inspiration. Once inspired, participants in a challenge may submit new ideas containing text and images, linking to existing ideas that inspired them. Over time, submitted ideas accrue views, applause, and comments as other participants provide feedback \cite{fuge:openideo_evolution_IDETC_2014}. Past work on helping filter ideas on OpenIDEO has investigated finding a small subset of diverse ideas \cite{ahmed:idetc_2016_idea} and ranking ideas by quality after training a classifier to identify winning ideas \cite{ahmed:cscw_2016_winning}.

We ran our experiment on 14 different challenges (domains) with total 3918 ideas submitted to these challenges. 
The challenge titles are shown in Table \ref{tab:table1}. To gain some intuition about how similar or different these domains are, Fig.~\ref{fig:tsne} projects the topic vectors of all ideas into 2-D using t-Distributed Stochastic Neighbor Embedding (t-SNE)~\cite{maaten2008visualizing}. 
t-SNE is a technique for dimensionality reduction that is particularly well suited for the visualization of high-dimensional datasets. The algorithm preferentially cares about preserving the local structure of the high-dimensional data. If two points are close in the original space, there is a strong attractive force between the points in the embedding, while if any two points are far apart in the original space, the algorithm is relatively free to place these points around.

As Fig.~\ref{fig:tsne} demonstrates, some challenges like challenge $14$ have many ideas which overlap with other challenges, while others like challenge $10$ have a tight cluster whose ideas largely differ from other challenges.
This disparity is expected; for example, challenge $10$ involves improving voting access during elections (a comparatively narrow and specific problem), while challenge $14$ addresses improving vibrancy in cities facing economic decline (a comparatively broad and open-ended problem). 
Later we find that this observation is further supported when we perform outlier detection, with some challenges having fewer outliers.
Topics in such challenges will not have high bisociation potential and are unlikely to be good bisociation candidates. This is because the bisociation score in Eq.~\ref{eq:btopic} is proportional to the number of outliers. Intuitively, if a domain is very narrow and specific (like bone marrow or voting challenges), it is less likely to gain from indirect connections with other domains.

Before demonstrating our model, the next section summarizes how existing state-of-the-art in BisoNet discovery\textemdash the ``CrossBee'' tool\footnote{\url{crossbee.ijs.si}}\textemdash performs on design examples to motivate the use of topics instead of words to bridge design domains. Thereafter, we discuss insights into the topic model and b-topics obtained using LDA. Then we run two experiments on Amazon's Mechanical Turk platform to gather human evaluation data. In the first experiment, workers assess whether the b-topics themselves are creative (compared to non-b-topics). In the second, we use the b-topics as inspiration in an idea generation task and ask workers to create or judge the generated ideas. Our experiments show that the above methodology is able to discover topics and produce ideas which people find creative.

\subsection{CrossBee Results: Comparison to existing state-of-the-art}
The existing state-of-the-art bisociation tool is the CrossBee (Cross Context Bisociation Explorer). It is an online tool to analyze text documents from two different domains. The tool finds and rank orders bridging terms (b-terms) but does not create a BisoNet. However, we demonstrate below that several issues arise when using words for design exploration rather than our proposed bridging topics. Since CrossBee can only handle two domains at once, we use it to find b-terms from Challenge $6$ (women's safety) and $14$ (city vibrancy) in our dataset as an illustrative and representative example. The full challenge topics are, respectively, ``How might we make low-income urban areas safer and more empowering for women and girls?''\footnote{\url{https://challenges.openideo.com/challenge/womens-safety/}} and ``How might we restore vibrancy in cities and regions facing economic decline?''\footnote{\url{https://challenges.openideo.com/challenge/vibrant-cities/}}

The top ten b-terms obtained using CrossBee between the women safety and city vibrancy challenge were: ``health, space, mobile, project, people, urban, community, city, program, area''. 
Without any gold standard data for b-terms on this particular example, it is difficult to say which of these b-terms are actually bisociative. However, looking at each term individually, one realizes that it is difficult to discover connections between these non-scientific domains by just using individual words like ``health'' or ``space''. Individual words like ``space'' can be ambiguous and may have different meanings depending on the context. Here ``space'' may refer to the space occupied by a body or related to the physical universe. However, a collection of semantically related words like ``space'', 
``outer'',``universe'', ``earth'',``atmosphere'' reduces ambiguity. This is unlike b-terms in autism-calcineurin dataset, where individual terms like ``paroxysmal'' or ``Bcl-2'' can lead one to discover links between two specific, scientific concepts because they are quite domain specific. Next, we contrast this with our method that incorporate our proposed b-topics rather than standard b-terms.

\subsection{Qualitative Results: Discovered B-Topics}
We run LDA with $100$ topics on all the $3918$ documents from $14$ domains and set the hyper-parameters for topic distribution smoothness and topic-word to values recommended in prior literature \cite{griffiths2004finding}. 
The output of LDA is topic-word and document-topic distributions along with the topics. To gain some intuition about LDA's output, we list the top 7 words for some of the learned topics:
\begin{enumerate}
\item food,	cook, meal,	recipe, restaurant, ingredient, eat
\item care,	health,	patient, hospital, doctor, medical,	center
\item money, bank, saving, funding,	pay, loan, financial
\item person, individual, need, van, match, contact, database
\end{enumerate}

These topics often (though not always, as shown by Topic 4) refer to some meaningful concept. Topic 1 above refers to food and restaurants, while Topic 2 refers to health care. 
Note that we have used LDA for topic analysis, but other topic model variants \cite{newman2011improving} can also be used.

To score these topics, we first find ideas that are outliers in a challenge. To train the classification model to classify ideas into challenges, the document-topic vectors were used as input. We trained multiple classification models to predict the domain, given vectors of ideas. For this dataset, three methods\textemdash Linear Discriminant, Bagged Trees, and Subspace discriminant \cite{MATLAB:2016}\textemdash had highest cross-validation accuracy of $87.5\%$, $88.2\%$ and $87.3\%$. The classification scores of these methods were added and the resultant method with $91\%$ accuracy was used to allocate predicted domains to every idea. The average F-1 score is $0.90$.
Ideas assigned to domains different from their true domain (false negatives) were identified as outliers. The number of outliers in each challenge is shown in Table~\ref{tab:table1}. 

Next, Eq.~\ref{eq:btopic} was used to find the topic bisociation score for every topic in every domain. To clarify and visualize our below explanations, we represent a topic by its top $10$ words, however, in reality each topic assigns a likelihood to every word and so other reasonable thresholding strategies could be used.

Let us take an example b-topic from challenge six (the womens' safety challenge): ``used, materials, design, clothing, wear, recycling, create, make, glass, shoe''. The idea with highest proportion of this b-topic is entitled ``Red Chilli Powder Filled Glass Bangle for Women's Self-defense.\footnote{\url{https://challenges.openideo.com/challenge/womens-safety/refinement/red-chilli-powder-glass-bangle-for-self-defence}}'' It discusses how a hollow glass bangle filled with hot red chili powder or pepper spray can be used by women in self-defense. 
This idea combines wearable accessories with self-defense mechanisms for women safety. The same b-topic has also been used in other contexts for different domains.
As one example, in challenge nine (related to unlawful detention of human rights activists), an idea entitled ``Emergency shoes'' proposes using special shoes with embedded wireless devices to help rights activists communicate their location to others in the event they are kidnapped or unlawfully detained. Multiple ideas across seemingly unrelated challenges\textemdash public safety, bone marrow registration, unlawful detention, among others\textemdash pursued a common theme of using clothing or wearable accessories as a possible solution. 
Surprisingly, this topic was the $5^{th}$ least used topic among all existing ideas, making the
concept quite rare. Such links may not be immediately obvious but once discovered can lead to different ideas than those that exist within the target domains.

As a second example, a different b-topic for challenge 6 (womens' safety) contains ``street, neighborhoods, residence, community, walk, tour, owners, home, local, house''. 
The topic relates to walking in neighborhoods and is the $3^{rd}$ least used topic. A representative idea from the women safety challenge for this topic is ``You'll Never Walk Alone\footnote{\url{https://challenges.openideo.com/challenge/womens-safety/ideas/walking-group}}'' which describes how women in low-income urban areas often share similar routes and could form walking groups by creating a group walking timetable between main points like bus stops. Likewise, in challenge 14 (city vibrancy) the idea entitled ``Youth Led Tours\footnote{\url{https://challenges.openideo.com/challenge/vibrant-cities/concepting/youth-led-tours}}'' proposes using local youths to guide visitors on walking tours through their cities, showing visitors the city as the residents see it. The women safety challenge took place three years after the vibrant city challenge, and many participants could arguably have gained insight from studying this related concept of combining a walking activity with womens' safety. However, most users were unlikely to have looked three years back in an unrelated challenge to discover such a connection. Using our method to mine bisociative links between seemingly unrelated domains can inspire people to propose such creative cross-domain solutions.

So far, we have discussed b-topics derived from domain outlier ideas, however, 
one can argue that creative links can also be found by using outlier topics directly, by identifying the most infrequently cited topics. 
However, we found this approach to be insufficient to identify bisociations, as these topics are often meaningless or completely unrelated to the problem in hand.
For example, for challenge six, we found the outlier topic to be: ``donation, donor, marrow, bone, registration, aware, register, people, swab, spread''. This topic, predominantly used in bone marrow related challenge 7, is completely unrelated to women safety and has practically zero proportion in current domain. Hence, adding such topics as exemplars does not identify a meaningful link.

Although, we have not studied the distance between domains directly, as a consequence of finding the b-topics among domain outliers, we find that the discovered bisociations are also from nearby domains, that is not from domains which are too ``far'' to share no outliers nor too ``near'' to be within domain.
In our analysis, the outliers are false negatives of the classification model, ideas which are far from their original domain, such that classifier confuses them to belong to another domain.
The b-topics are topics which are common between outliers of two domains and uncommon overall. This generally means these topics are on concepts which are far from the mainstream concepts of the domain, but not very far from the domain to be absent from the outlier. As the bisociation score of a topic is proportional to the number of outliers (Eq.~\ref{eq:btopic}), domains with more outliers (hence more nearby domains) have higher chances of discovering true bisociations. For example, the voting challenge has only seven outliers, hence topics get a low bisociation score in it, implying that it is unlikely to find an indirect connection from other domains. The challenge is narrow in scope and far away from all other domains (as visualized in Fig.~\ref{fig:tsne}).
Further research is needed to establish if discovering bisociations using outlier method supports previous research in \cite{fu2013meaning,chan2014best} showing that the ``distance" away from the design problem of the creativity stimulus has an influence on the quality of the new solution.

\subsection{Qualitative Results: Exploring the Resulting BisoNet}

So far, we discussed ranking topics by their bisociation potential.
Next, we create a BisoNet across challenges to explore concepts which can be borrowed between challenges. 
Note that for 14 challenges, if existing word-based BisoNets without pruning are used with a global vocabulary size of $2000$, the number of nodes in the network will be $28,000$. This will make graph exploration difficult, if not impossible. By using $100$ topics, we reduce the network size by $95\%$ to $1400$ nodes. However, to further help network exploration, one can optionally obtain further reduction by node removal and edge pruning methods. 

\begin{figure}
\includegraphics[width=\columnwidth]{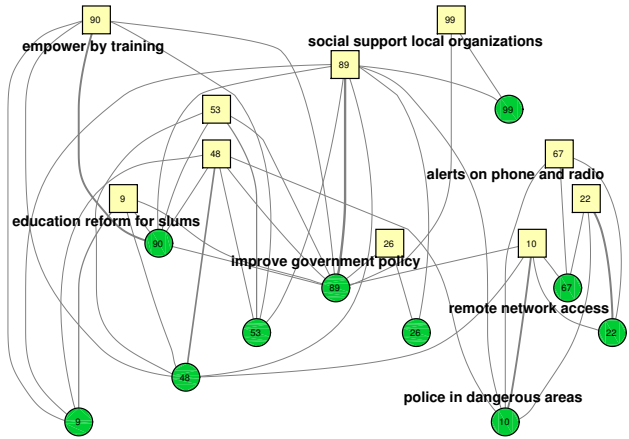}
\caption{A snapshot of BisoNet showing links between topics between challenge six and nine addressing, respectively, womens' safety and gathering information from hard to access areas. We only show largest connected component after thresholding to top $0.5\%$ edges with highest bison similarity. Node with id $6\_9$ represents challenge six with topic id nine. Higher edge weights are shown with thicker lines. Major themes of the topics are captioned. }
\label{fig:bisonet}
\end{figure}

Formatting the full BisoNet of all $14$ challenges legibly in this paper is difficult, so for clarity we discuss and visualize a smaller sample.
Figure~\ref{fig:bisonet} shows a small subset of a full BisoNet by viewing the portion connecting challenge six and nine. Challenge six addresses women safety and empowerment while challenge nine addresses gathering information from hard to access areas to prevent mass violence. To make network visualization easy, we only show the largest connected component of the graph after retaining the edges with the top $0.5\%$ of edge weights and use parameter value $k=0.5$ in Eq.~\ref{eq:2}. This BisoNet has $20$ nodes, representing ten topics. Challenge nine is shown by yellow squares, while challenge six is shown by green circles.
A larger edge weight (thicker line) between two different topics mean that topics may co-occur together in similar proportions in same ideas in these challenges. A larger edge weight between the same topic across two challenges illustrates that it has a high topic bisociation score (Eq.~\ref{eq:2}).


By inspecting the graph in Figure~\ref{fig:bisonet}, we find topics that likely refer to a few broad themes that can apply to both domains. For example, the right side of the graph has topics on technology related solutions like network coverage or phone messaging with top words in topics being ``device, use, technology, area, signal'' (Topic 22) and ``phone, message, send, text, mobile'' (Topic 67). Within challenge 9, we find that the idea with highest proportion of latter topic is entitled  ``Balloon Communications''\textemdash it proposes flying an iridium based sat-phone as a weather balloon payload over the affected area and receive/transmit text messages from local cell phones.
Another idea in challenge nine proposed a text message based wristband that can send any number of predefined messages to a connector, network, or hub. Related to same topic in challenge six on womens' safety, we find linked ideas like creating a mobile application that can deter assault by automatically notifying your emergency contacts if the user does not travel from their stated start and end points safely or quickly. By using the proposed BisoNet to isolate these concepts that share b-topics across different challenges, we could promote more effective cross-pollination of ideas.

Similarly, in Figure~\ref{fig:bisonet} we find the left side topics are related to education and training (Topic 9 with words ``girl, community, slums, schools'' and Topic 90 with words ``woman, income, training, urban''). These are predominant in challenge 10 (womens' safety) and not heavily used in challenge 9. Uncovered themes include government policy improvements and community support. For example, within community support,
a challenge 9 idea entitled ``Reflexive distributive community warning system'' talks about organizing channels of communication and introducing universal codes that could increase speed of transmission and accuracy of information within a community. It mentions steps to design codes for danger, layout the location of each community in the area, and instructs each village with specific actions to undertake if they experience or witness atrocities. A similar concept of community preparation and action could likely also apply to safety in urban areas. Within the womens' safety challenge, one idea possessing this b-topic talks about establishing community-fitness centers to create a larger network of people who can recognize each other on the streets, commute together, and feel an overall sense of community. This idea discusses building a community, while the idea in challenge nine addresses action after the community is formed. Through the process of BisoNet graph exploration and use of b-topics to guide inspiration, our hypothesis is that ideas and concepts from other domains can help designer better explore, cross-pollinate, or gain inspiration within their own domain. As stated by Pioncar{\'e} above, while not all links may be useful, some may give valuable insights.

\subsection{Quantitative Results: Human Evaluation}
Our subjective analysis demonstrated a subset of useful b-topics and cross-domain links that the proposed BisoNet method identified. 
However, verifying BisoNet performance directly and objectively is difficult, as creative inspiration depends on human perception and there is no accepted gold-standard dataset within design (unlike those for existing word-based BisoNets using in academic literature search~\cite{schmidt2012bisociative}). Moreover, quantitatively comparing our topic based links with those of the word-based Crossbee b-terms would not represent a fair comparison, since our b-topics contain strictly more information compared to a single b-word.

One possible baseline against which to compare our method is to create topics using LDA, but, rather than going through the effort of finding b-topics, just show a designer a random topic from LDA as inspiration and compare the outcomes. However, this may be a comparatively weak baseline, as topics produced by LDA can vary in coherence and human interpretability. To create a stronger baseline, we 
calculate Pointwise Mutual Information (PMI) for each topic and pick a random topic with similar PMI.
Recent work \cite{newman2010visualizing} has shown that PMI can be used to estimate human-judged topic coherence\textemdash hence the baseline random topic (r-topic) is similar in coherence to b-topic, resulting in a fairer comparison. To measure topic coherence, we use normalized PMI score, calculated over the entire Wikipedia corpus. 

To compare the creativity of b-topics with an r-topic, we use crowd workers on Amazon's Mechanical Turk platform. One na{\"i}ve way to quantitatively compare b- and r-topics topic is to find existing ideas from the challenges with a high proportion of a b-topic versus r-topic and ask workers to rate the idea on quality and novelty. Although straightforward to implement, results from such an approach may be misleading. Two ideas on OpenIDEO may differ for multiple reasons\textemdash poor grammar, domain knowledge of author, \etc~Thus, workers should ideally compare topics or ideas generated by the same author, where the only difference lies in the seed topic used for inspiration. To address this, we conducted two randomized experiments to answer two research questions:
\begin{itemize}
\item Are b-topics perceived as creative?
\item Do b-topics, when used for creative inspiration, produce more creative ideas?
\end{itemize}

In both of our subsequent experiments, we use crowd-sourcing to both generate and evaluate the creativity of the generated ideas, building upon techniques used by researchers in both engineering and computer supported collaborative work \cite{green2014crowd,Kittur:2013:FCW:2441776.2441923,kittur2010crowdsourcing}.

\subsubsection{Quantitative Experiment 1: Are b-topics perceived as creative?}

\begin{figure}
\includegraphics[width=\columnwidth]{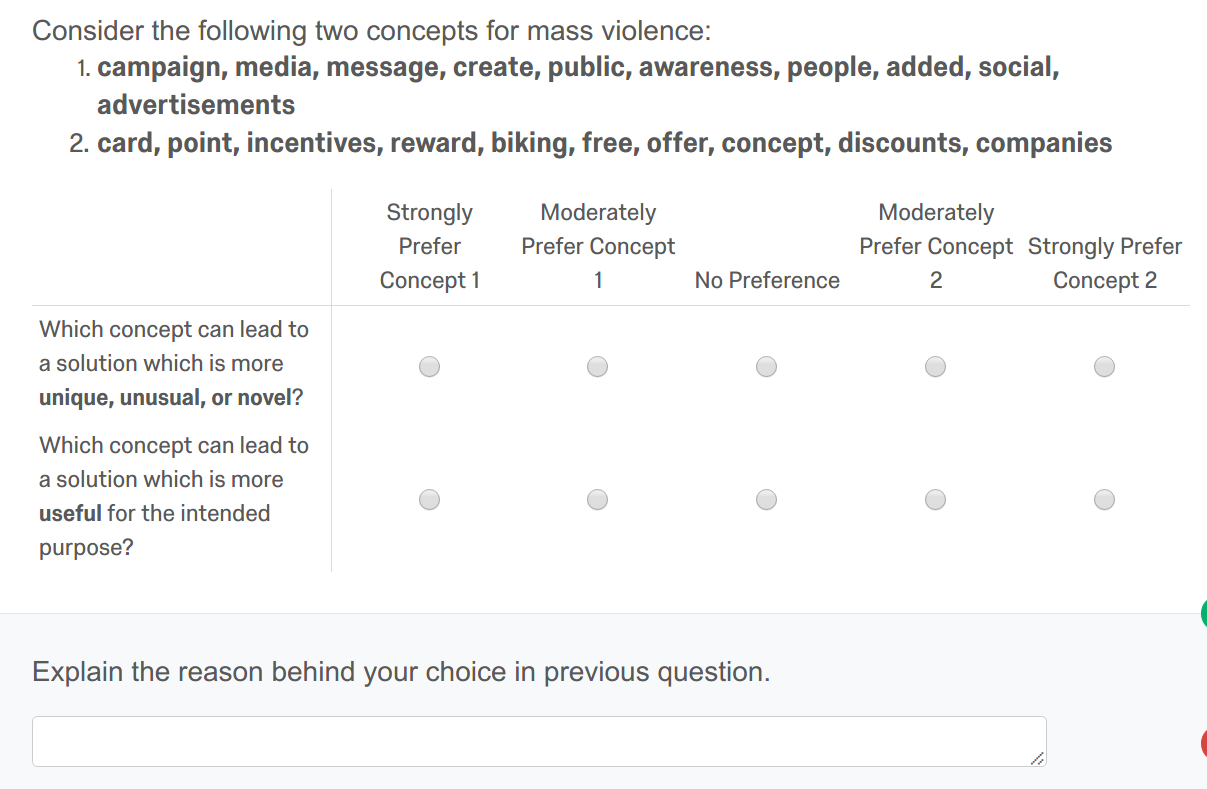}
\caption{Objective survey example}
\label{fig:survey1}
\end{figure}

\begin{figure}
\includegraphics[width=\columnwidth]{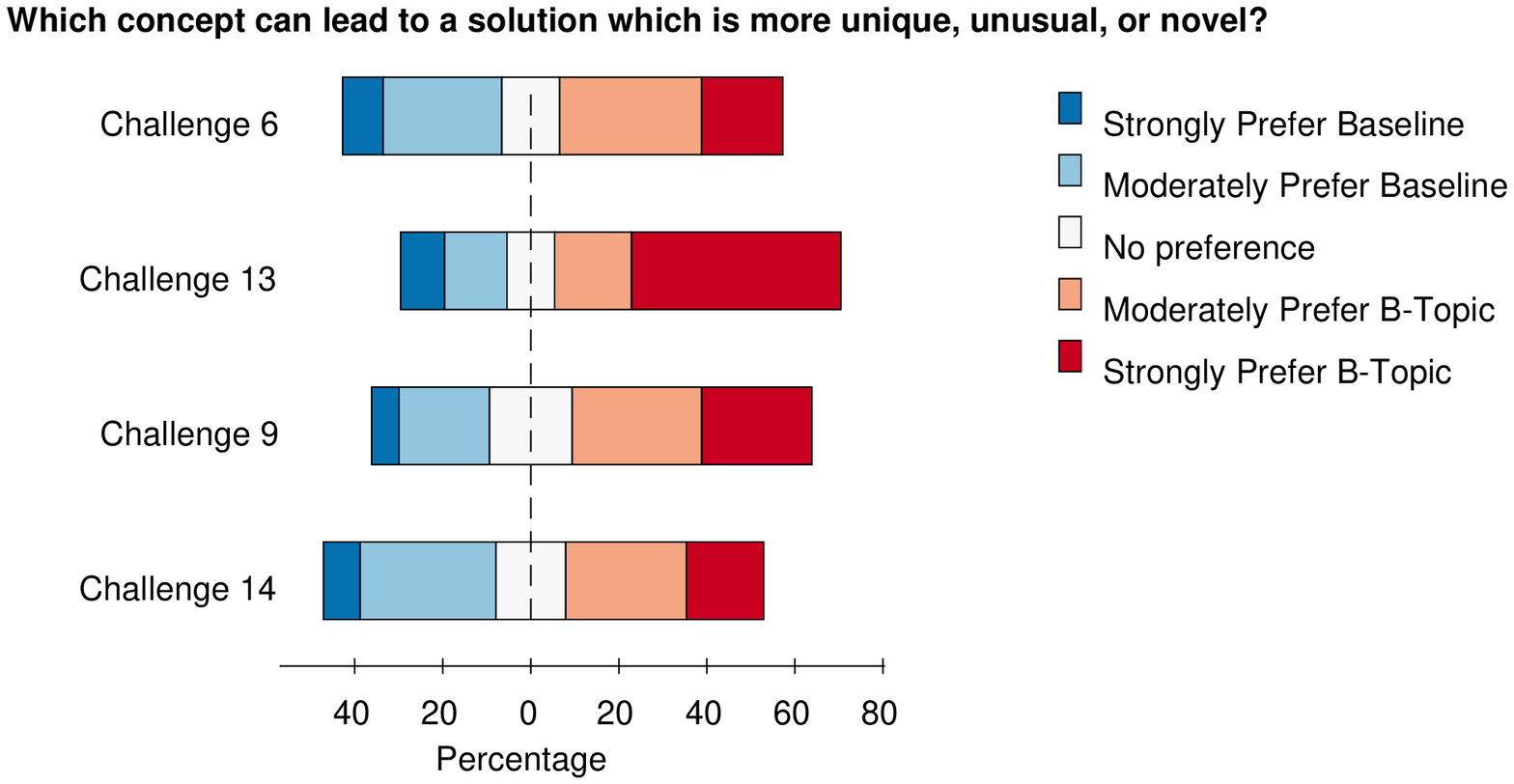}
\caption{Novelty scores from objective assessment. Each challenge had 4 B-topic comparisons which were rated by 30 workers.}
\label{fig:on}
\end{figure}

\begin{figure}

\includegraphics[width=\columnwidth]{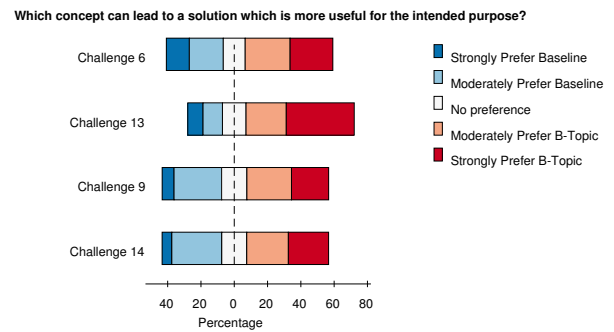}
\caption{Quality scores from objective assessment. Each challenge had 4 B-topic comparisons which were rated by 30 workers.}
\label{fig:oq}
\end{figure}

Here we consider the hypothesis that b-topics can find more creative links between design domains compared to baseline. To test this, we first showed both b-topics and baseline topics to online workers and asked them to rate the topics themselves. 
To design the survey, we selected four challenges and four topics in each challenge. 
Four b-topics were selected randomly from the top ten topics with highest bisociation score for a challenge (total 16 b-topics for four challenges.)
While we chose the challenges randomly, we did not select a challenge with very few outliers, as all the topics in such challenges have low bisociation scores.

For the baseline, we found three random topics that had similar PMI scores to the b-topic and selected the one with the lowest bisociation score (the random baseline topic should not also be a b-topic). We showed the challenge brief to 30 Turkers for each pair of b-topic and r-topic. For novelty, the workers were asked ``which topic can lead to a solution which is more unique,'' while for quality, the workers were asked ``which concept can lead to a solution which is more useful for the intended purpose'' as shown in Fig.~\ref{fig:survey1}. These  survey questions are based on \cite{pang2016crowdsourcing}, where novelty and quality questions were used to find concepts which are more creative. We ensured that the crowdsourced responses were valid using a few quality checks. 
First, we
allowed only those Turkers to participate, whose acceptance rate of past work was more than $95\%$.
Second, in every survey, we added one subjective question asking Turker to explain the rationale behind their choice. 
Some Turkers, who were only trying to maximize questions answered (and thus not meaningfully participating), often entered gibberish to this question and their responses were discarded. 
Finally, we also recorded time at task and number of clicks on page to filter out participants whose metrics were obvious outliers.

The survey results are shown in Fig.~\ref{fig:on} and Fig.~\ref{fig:oq}, 
where we notice that most workers preferred b-topics for both novelty and quality compared to other topics. 
We do not report statistical significant analyses for these experiments, as the Likert scales are ordinal and comparison for a particular domain are between different sets of topics with varying b-scores.
In some assessments, the workers were asked to explain their choice, giving us useful insights into their thinking. For example, one user who strongly preferred concept 2 for novelty but prefers concept 1 for quality in Fig.~\ref{fig:survey1} quotes ``I have never heard of discounts or offers or free things as a reward to help inform about mass violence. Advertisements, social interaction, campaigns and interacting with the public seem more useful, as many victims involved in areas where such crimes take place are not interested in discounts or free offers to stop violence. They just want the violence stopped.'' while another user who supports the b-topic says ``Concept 1 sounds like 'getting the word out' about atrocities and therefore hoping to prevent them. Concept 2 sounds like a tacky marketing ploy.''

\paragraph{Experiment 1 Limitations:} 
This experiment was a more direct way of measuring perceived novelty and quality of generated b-topics. Although our results showed that b-topics are perceived to be more creative than the  baseline, this observation should be taken with a grain of salt. First, we represent a topic by its top 10 words for sake of clarity. However, it is possible that discarding those lower-ranked words might subtly change the topic's perceived meaning. Second, we presented the words in order of their proportion in the topic. It is possible that using a word cloud or a different ordering of the same words within the top 10 may alter a topic's perception by the online workers.\footnote{To test this effect, we jumbled the words of a topic to generate two collection of words\textemdash ``production, market, selling, customer, increase, produce, brand, supply, sale, distribution'' and ``produce, supply, distribution, market, customer, sale, brand, production, increase, selling''. These two groups use the same set of 10 words and were given to workers to compare novelty and quality between them. Only $50\%$ of the workers gave no preference between the topics. This observation could either mean lack of attention on task or the concept perceived by a worker from reading the words in a topic is dependent on word ordering.} Studying both effects would be an interesting area of future work but is not further addressed in this paper. 

Another limitation of this experiment is that it is not straightforward to run direct tests of statistical significance to test whether or not b-topics are perceived as more creative than the baseline, in part because the differences in bisociation scores come from different populations and effect sizes, complicating traditional inference models. 
Despite these limitations, we studied within category (e.g., Prefer B-Topic, Neutral, etc.) trends for each challenge, as the difference in bisociation score between a random topic and b-topic increases. To do so, we noted differences in the b-scores between a b-topic and random topic with respect to the proportion of the response rate from the survey participants.
Ideally, increasing difference in bisociation should lead to stronger relative preference for b-topics over random topics. We find that,
for each challenge, a higher percentage of respondents preferred b-topics over the baseline and Challenges 6, 9, and 14 mirror the slope behavior we would expect while challenge 13 does not. 
However, as the slope estimates are noisy and fairly small in magnitude, it is difficult to make strong statements about the effect. 
As such, our results should be interpreted with appropriate caution.

\subsubsection{Quantitative Experiment 2: Do b-topics produce more creative ideas?}

\begin{figure}
\includegraphics[width=\columnwidth]{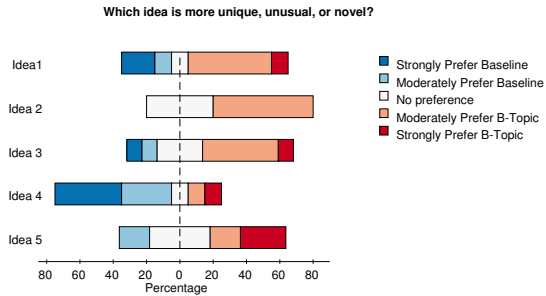}
\caption{Novelty scores for ideas on topic ``city, local, government, create, need, people, urban, citizens, economic, new'' vs ``garden, growing, farming, urban, plant, food, land, vegetables, community, fruits
 ''. Each idea pair is rated by 10 workers.}
\label{fig:subjn}
\end{figure}

\begin{figure}
\includegraphics[width=\columnwidth]{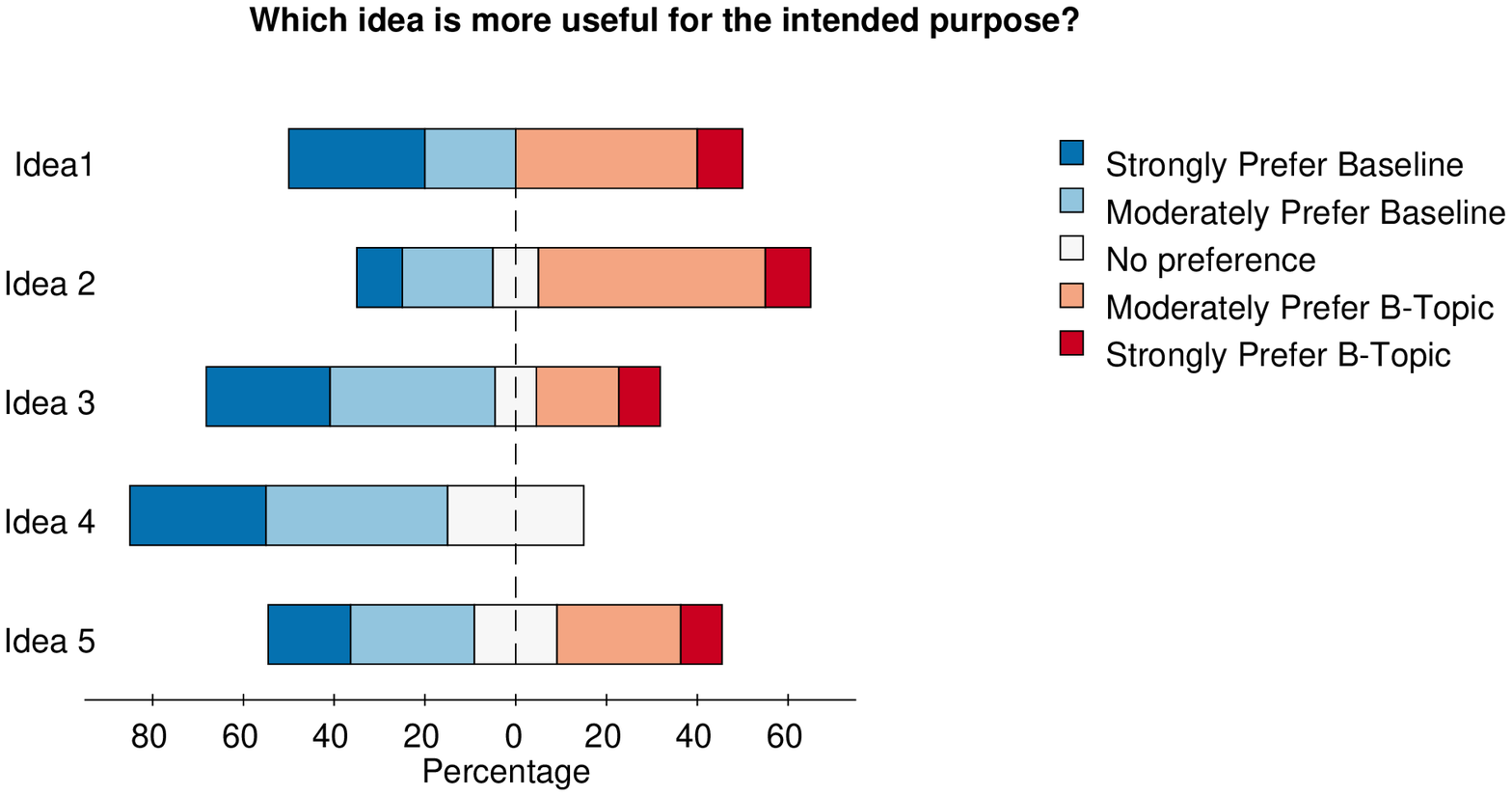}
\caption{Quality scores for ideas on topic ``city, local, government, create, need, people, urban, citizens, economic, new'' vs ``garden, growing, farming, urban, plant, food, land, vegetables, community, fruits''. Each idea pair is rated by 10 workers.}
\label{fig:subjq}
\end{figure}

\begin{table}
\centering
\begin{raggedleft}
\begin{scriptsize}
\begin{tabular}
{|p{1cm}|p{3.1cm}|p{2.5cm}|}
\hline
Challenge  & Baseline Topic: `city, local, government, create, need, people, urban, citizens, economic, new'  & B-Topic: `garden, growing, farming, urban, plant, food, land, vegetables, community, fruits' \\
\hline
How might we restore vibrancy in cities and regions facing economic decline? & I feel that citizens within an urban environment need to work with city and local government to identify and create new economic programs and opportunities to make their locations vibrant. Within the city, the main motivation will have to be for the citizens to want to improve their lives and surroundings. It starts from within by showing pride in their neighborhoods. I think that it starts small with something like a community garden in which a vacant lot or piece of land is turned into something that can be a positive for the community as a whole. Street beautification and just an overall caring about the neighborhood can lead to a turn around but it starts with the citizens needing to want it to happen.  & I feel that since most urban neighborhoods do not have supermarkets that are convenient, it might be a huge plus to start urban farming. A community garden in which fruits and vegetables can be grown and sold to the public who have no other access to them. I think that vacant or under used land would be a wonderful place to put this. I think the citizens can plant and grow the vegetables and fruits and sell them which would make the project economically feasible as well. It would be self sustaining and it would be a source of pride for the people.  \\
\hline
\end{tabular}
\caption{Sample ideas submitted by a crowd worker on two topics}
\label{tab:city}
\end{scriptsize}
\end{raggedleft}
\end{table}

\begin{figure}
\includegraphics[width=\columnwidth]{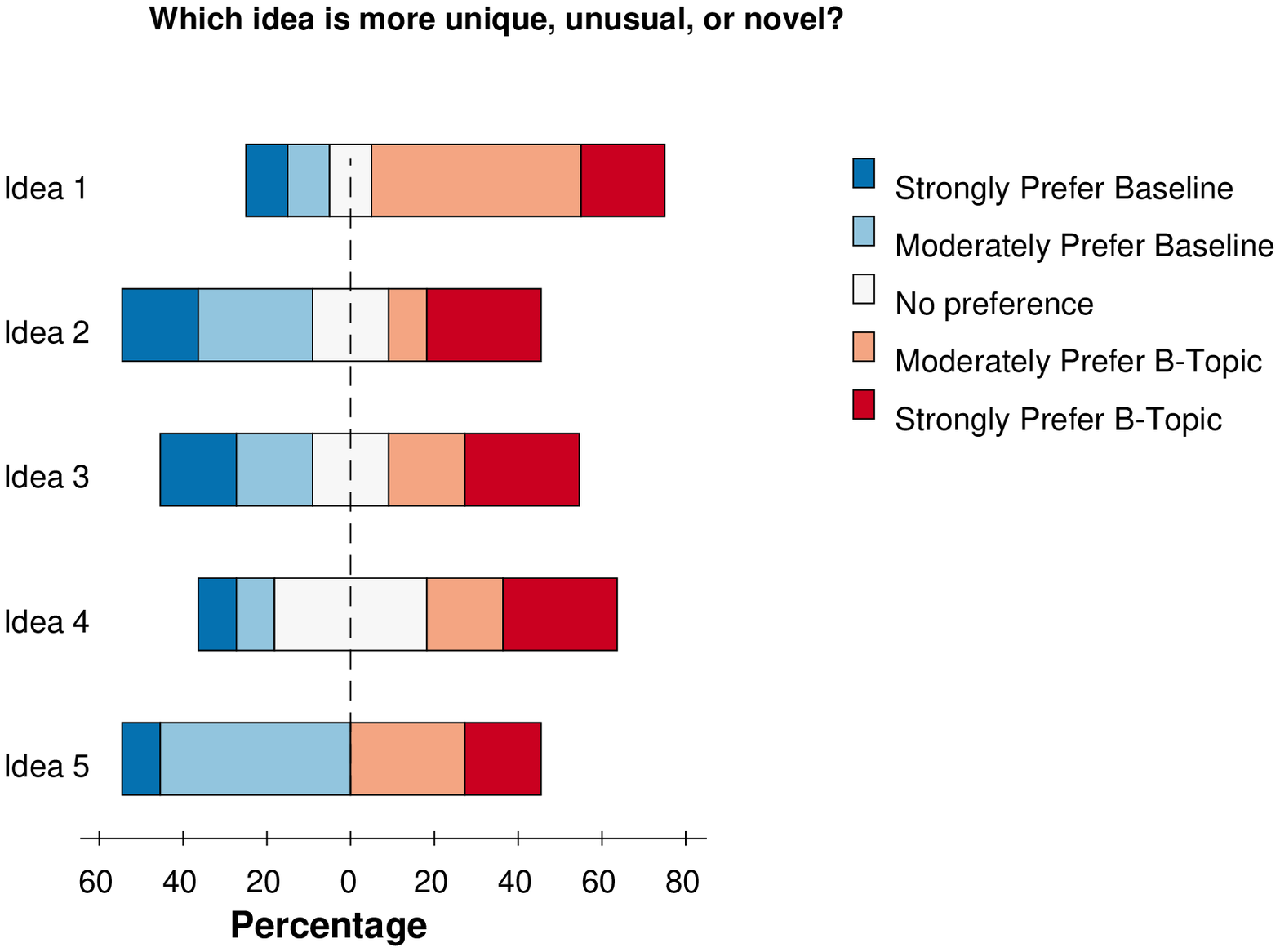}
\caption{Novelty scores for ideas on topic ``woman, safety, safe, areas, urban, community, low, city, ideas, income'' vs ``device, use, technology, area, signal, network, community, access, people, remote''. Each idea pair is rated by 10 workers.}
\label{fig:subjn_women}
\end{figure}

\begin{figure}
\includegraphics[width=\columnwidth]{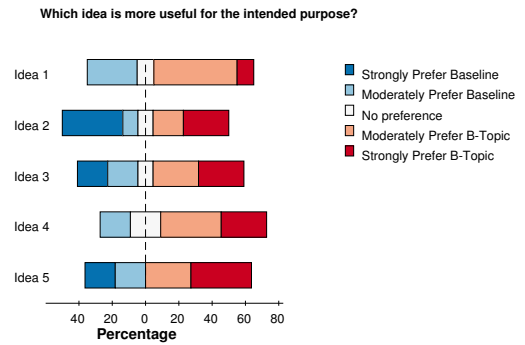}
\caption{Quality scores for ideas on topic ``woman, safety, safe, areas, urban, community, low, city, ideas, income'' vs ``device, use, technology, area, signal, network, community, access, people, remote''. Each idea pair is rated by 10 workers.}
\label{fig:subjq_women}
\end{figure}

Next, we tested whether b-topics inspire more creative ideas by conducting a set of idea generation experiments and evaluations. First, we provided a few workers with the challenge brief and two topics. The workers were asked to write an original solution to the problem in more than 100 words. They were explicitly instructed to use the set of words from the provided topic (collection of 10 words) as inspirations to their idea. 

Each worker was asked to perform this task twice, using two different topics: a b-topic and, as a baseline, the topic with highest proportion for the challenge prompt they received (most common topic). 
For a given pair of topics, we generate five pair of ideas from 5 workers. The workers are asked to self assess their ideas on quality and novelty. Next, we judge the quality and novelty of these idea pairs (ideas generated by same worker) by asking another, independent set of 10 workers to compare these ideas on quality and novelty. The order of ideas is randomized and to remove possible bias on novelty, we do not repeat judges, hence using 50 unique workers. 
The experiment was done on Challenge $14$ on improving city vibrancy and the topics are shown in Table~\ref{tab:city}. Figures~\ref{fig:subjn} and~\ref{fig:subjq} show the quality and novelty results, respectively. 

Overall, we found that the workers judged ideas generated using the b-topic as more novel but not necessarily higher quality.
When the writers of an idea were asked to rate their own ideas\textemdash \ie, the one generated with the b-topic prompt versus the baseline common topic\textemdash for novelty, three writers gave no preference while one strongly preferred the b-topic idea and one moderately preferred the b-topic idea.
For quality, one strongly preferred the b-topic idea, two writers moderately preferred the b-topic idea, one gave no preference and one strongly preferred the baseline idea.
Table~\ref{tab:city} compares a sample idea pair. When the idea was evaluated by the independent raters, the b-topic idea received more favorable ratings for both novelty and quality, as shown in Fig.~\ref{fig:subjn} and Fig.~\ref{fig:subjq}. 

We have two main observations. First, the idea writers were able to draw a connection between seemingly unrelated topic on 'garden, growing, farming' and propose novel ideas on city vibrancy. Secondly, the raters found the ideas prompted by b-topics more novel. Perhaps not surprisingly, the challenge topic was found to be more useful, compared to the b-topic, in part because it directly addressed the challenge issue. 

To further test, how the results generalize to other challenges, we conducted the same experiment for Challenge $6$ on women safety. Here, the b-topic was ``device, use, technology, area, signal, network, community, access, people, remote" and the most common topic for this challenge was ``woman, safety, safe, area, urban, community, low, city, ideas, income". We found that the b-topic was rarely used in this challenge, with only eight ideas having it as the highest proportion topic. Fig.~\ref{fig:subjn_women} and Fig.~\ref{fig:subjq_women} show the novelty and quality scores for set of five idea pairs generated by five Turkers and rated by another ten each. We find, that for this challenge too, b-topic was preferred for both novelty and quality.

\paragraph{Experiment 2 Limitations:} 

While this experiment tested how useful b-topics were for inspiring creative ideas, it comes with a few caveats. First, it is difficult to guarantee how much of the idea was inspired by the novel connection provided by the topic; \ie, we could not force them to use the topic, though, anecdotally, by and large the ideas did appear to leverage the provided topic. In addition, there can be variations within the quality of work that the workers produce due to a writer's past knowledge in a domain. 
Likewise, novice or non-imaginative writers may fail to see a relationship or connection between the challenge theme and b-topic, compared to the more obvious connections with the baseline common topic that is closer to the challenge domain. 
We also cannot isolate a particular topic; \ie, a generated idea may well use other concepts too, so the final ratings of an idea can depend on multiple factors beyond the chosen b-topic. Lastly, given that this experiment consists of 50 evaluations by raters from Amazon's Mechanical Turk service, we should be careful when generalizing these results other domains and rater populations; replicating these results with additional experiments on alternative rater populations would provide useful comparisons.
The monetary motivation and the time constraints for Turkers may also impact the experimental results. Lastly, as with experiment 1, it is not straightforward to run direct tests of statistical significance to test whether or not b-topics statistically more creative ideas than the baseline, again due to differences in bisociation scores from different populations with complicates traditional statistical inference models. As such, our results should be interpreted with appropriate caution.

\section{Conclusions, Limitations, and Future Work}
This paper presented a method for exploring cross-domain design ideas through the use of Bisociative Information Networks (BisoNets). Specifically, it introduced the use of bridging topics (b-topics) and generalized past results in BisoNets to allow simultaneous exploration of multiple domains. 
The paper demonstrated this capability on an example of design exploration and discovery using a dataset of thousands of ideas from OpenIDEO, an online collaborative community. In doing so, it answered the following two questions 1) Are b-topics perceived as creative? 2) Do b-topics, when used for creative inspiration, produce more creative ideas?

Our qualitative results demonstrated the limitations of existing BisoNet techniques when applied to non-specialist domains, along with the usefulness of representing conceptual bridges through collections of words (topics) rather than single terms. We also demonstrated the usefulness and efficiency of finding cross-domain inspiration from collections of thousands of ideas; such techniques have direct applications for both large-scale design ideation, in addition to traditional design search and retrieval for analysis of patents or other analogical stimuli.
Our quantitative results demonstrated that b-topics, whether presented on their own or via ideas they inspired, were generally viewed as more novel, though not necessarily higher quality, compared to non-b-topic baselines. We also found that b-topics, when used for creative inspiration, helped produce more creative ideas compared to most common topic for a domain.

These findings show that bridging concepts can be found in outlier ideas which belong to one domain, but are confused to belong to another. 
Due to the rarity of these outlier ideas in the current domain, such links may not be immediately obvious, but once discovered they can lead to creative ideas. 
In contrast to past work, we show 
that representing the bridging concepts using latent topics is advantageous over single words.
We also differ from past approaches which use distance metrics by using a classification model for outlier detection. This has the inherent advantage of finding bisociations depending on the distribution of ideas between domains and not distance between them.
These outlier ideas help identify bisociations far from the mainstream concept, but not very far from the domain. 

The main limitations of our proposed techniques are two-fold. First, our method relies on generating good topic distributions for each idea. 
With the available multitude of topic model variants, this is easier said than done. 
We used standard LDA to find collections of words organized in a fixed number of topics. In an unknown domain, it is difficult to know how many topics exist (though there are non-parametric, countably infinite dimensional LDA variants that can handle this~\cite{teh2004sharing}).
As topic models themselves are not aware of existing bisociations; an interesting albeit challenging area of research would be to incorporate bisociation principles within the LDA update equations, so that topics found are more likely to be bisociative. 

Second, even if the b-topics themselves are accurate, they still require some creative imagination on the part of the designer to connect the b-topic to the challenge at hand. While presenting collections of words or exemplar ideas are two straightforward mechanisms to help spark this inspiration, future research could address the open question as to what format or intervention would best help designers internalize or connect ideas across domains.
It would also be interesting to study 
the effect of topics with similar bisociation score but varying in other attributes like the degree of abstraction and the type of words used.
One of the directions of future work can be to study effectiveness of b-topic when it is dominated by certain type of words like functional words or behavior words.

Lastly, the use of topics or word collections as a vehicle to bridge two domains is a somewhat blunt (if effective) instrument, when compared to other more structured analogical reasoning approaches that require more detail about a design idea beyond just unstructured text\textemdash \eg, Gentner's Structure-Mapping framework \cite{gentner1983structure} or the use of Functional Basis Structures in biologically inspired design~\cite{cheong2011biologically}. Merging bisociation with those more formal analogical structures could provide the best of both worlds.

\section{Acknowledgements}
This work was partially funded through a University of Maryland Minta Martin grant, as well as support from the Department of Mechanical Engineering. Partial funding for open access provided by the UMD Libraries' Open Access Publishing Fund.

\begin{table}[ht]
\centering
\begin{tabular}{|p{7cm}|p{1.4cm}|p{1.6cm}|}
\hline
\textbf{Title} & \textbf{Number of ideas} & \textbf{Number of outliers}\\
\hline

1. How might we better connect food production and consumption? & 606 & 45\\

2. How can technology help people working to uphold human rights in the face of unlawful detention? & 165 & 22\\

3. How might we support web entrepreneurs in launching and growing sustainable global businesses? & 157 & 17\\

4. How can we raise kids awareness of the benefits of fresh food so they can make better choices? & 180 & 16\\

5. How might we create healthy communities within and beyond the workplace? & 240 & 12\\

6. How might we make low-income urban areas safer and more empowering for women and girls? & 573 & 50\\

7. How might we increase the number of registered bone marrow donors to help save more lives? & 285 & 11\\ 

8. How might we improve maternal health with mobile technologies for low-income countries? & 176 & 12 \\

9. How might we gather information from hard-to-access areas to prevent mass violence against civilians? & 166 & 27\\

10. How might we design an accessible election experience for everyone? & 154 & 7\\

11. How might we all maintain well-being and thrive as we age? & 134 & 13\\

12. How can we equip young people with the skills, information and opportunities to succeed in the world of work? & 148 & 35\\
13. How might we inspire young people to cultivate their creative confidence? & 608 & 41\\
14. How might we restore vibrancy in cities and regions facing economic decline? & 326 & 48\\

\hline
\end{tabular}
\caption{14 Challenges incorporated in dataset showing the size of the challenge and number of outliers}
\label{tab:table1}

\end{table}

\bibliographystyle{apalike}
\bibliography{sample}

\begin{thebibliography}{}

\bibitem[Ahmed et~al., 2016]{ahmed:idetc_2016_idea}
Ahmed, F., David~Gorbunov, L., and Fuge, M. (2016).
\newblock Discovering diverse, high quality design ideas from a large corpus.
\newblock In {\em Proceedings of ASME International Design Engineering
  Technical Conferences (IDETC'16)}. ASME.

\bibitem[Ahmed and Fuge, 2017]{ahmed:cscw_2016_winning}
Ahmed, F. and Fuge, M. (2017).
\newblock Capturing winning ideas in online design communities.
\newblock In {\em Computer-Supported Cooperative Work and Social Computing
  (CSCW 2017)}.

\bibitem[Benami and Jin, 2002]{benami2002creative}
Benami, O. and Jin, Y. (2002).
\newblock Creative stimulation in conceptual design.
\newblock {\em Proceedings of ASME DETC/CIE, Montreal, QC, Canada,
  DETC2002/DTM-34023}, 2(1).

\bibitem[Berthold, 2012]{berthold2012towards}
Berthold, M.~R. (2012).
\newblock Towards bisociative knowledge discovery.
\newblock In {\em Bisociative Knowledge Discovery}, pages 1--10. Springer.

\bibitem[Blei and Lafferty, 2007]{blei2007correlated}
Blei, D.~M. and Lafferty, J.~D. (2007).
\newblock A correlated topic model of science.
\newblock {\em The Annals of Applied Statistics}, pages 17--35.

\bibitem[Blei and Lafferty, 2009]{blei2009topic}
Blei, D.~M. and Lafferty, J.~D. (2009).
\newblock Topic models.
\newblock {\em Text Mining: Classification, Clustering, and Applications},
  10(71):34.

\bibitem[Blei et~al., 2003]{blei2003latent}
Blei, D.~M., Ng, A.~Y., and Jordan, M.~I. (2003).
\newblock Latent dirichlet allocation.
\newblock {\em the Journal of Machine Learning Research}, 3:993--1022.

\bibitem[Boden, 1994]{boden1994precis}
Boden, M.~A. (1994).
\newblock Pr{\'e}cis of the creative mind: Myths and mechanisms.
\newblock {\em Behavioral and brain sciences}, 17(3):519--531.

\bibitem[Chan et~al., 2015]{chan2014best}
Chan, J., Dow, S.~P., and Schunn, C.~D. (2015).
\newblock Do the best design ideas (really) come from conceptually distant
  sources of inspiration?
\newblock {\em Design Studies}, 36:31--58.

\bibitem[Chan et~al., 2014]{chan2014overreliance}
Chan, J., Schunn, C., and Dow, S. (2014).
\newblock Overreliance on conceptually far sources decreases the creativity of
  ideas.
\newblock In {\em Proceedings of the Cognitive Science Society}, volume~36.

\bibitem[Chan and Schunn, 2015]{chan2015importance}
Chan, J. and Schunn, C.~D. (2015).
\newblock The importance of iteration in creative conceptual combination.
\newblock {\em Cognition}, 145:104--115.

\bibitem[Chang and Blei, 2009]{chang2009relational}
Chang, J. and Blei, D.~M. (2009).
\newblock Relational topic models for document networks.
\newblock In {\em International Conference on Artificial Intelligence and
  Statistics}, pages 81--88.

\bibitem[Cheong et~al., 2011]{cheong2011biologically}
Cheong, H., Chiu, I., Shu, L., Stone, R., and McAdams, D. (2011).
\newblock Biologically meaningful keywords for functional terms of the
  functional basis.
\newblock {\em Journal of Mechanical Design}, 133(2):021007.

\bibitem[Cilibrasi and Vitanyi, 2007]{cilibrasi2007google}
Cilibrasi, R.~L. and Vitanyi, P.~M. (2007).
\newblock The google similarity distance.
\newblock {\em IEEE Transactions on knowledge and data engineering}, 19(3).

\bibitem[DeYoung and Hobbs, 2009]{deyoung2009discovery}
DeYoung, D. and Hobbs, D. (2009).
\newblock {\em Discovery of Design: Searching Out the Creator's Secrets}.
\newblock New Leaf Publishing Group.

\bibitem[Dubitzky et~al., 2012]{dubitzky2012towards}
Dubitzky, W., K{\"o}tter, T., Schmidt, O., and Berthold, M.~R. (2012).
\newblock Towards creative information exploration based on koestler’s
  concept of bisociation.
\newblock pages 11--32.

\bibitem[Fu et~al., 2013]{fu2013meaning}
Fu, K., Chan, J., Cagan, J., Kotovsky, K., Schunn, C., and Wood, K. (2013).
\newblock The meaning of “near” and “far”: the impact of structuring
  design databases and the effect of distance of analogy on design output.
\newblock {\em Journal of Mechanical Design}, 135(2):021007.

\bibitem[Fuge and Agogino, 2014]{fuge:openideo_evolution_IDETC_2014}
Fuge, M. and Agogino, A. (2014).
\newblock How online design communities evolve over time: the birth and growth
  of {OpenIDEO}.
\newblock In {\em ASME 2014 International Design Engineering Technical
  Conferences and Computers and Information in Engineering Conference}.
  American Society of Mechanical Engineers.

\bibitem[Fuge et~al., 2014]{fuge:openideo_JCISE_2014}
Fuge, M., Tee, K., Agogino, A., and Maton, N. (2014).
\newblock Analysis of collaborative design networks: A case study of
  {OpenIDEO}.
\newblock {\em Journal of Computing and Information Science in Engineering},
  14(2):021009+.

\bibitem[Gentner, 1983]{gentner1983structure}
Gentner, D. (1983).
\newblock Structure-mapping: A theoretical framework for analogy.
\newblock {\em Cognitive science}, 7(2):155--170.

\bibitem[Gon\c{c}alves et~al., 2012]{gonalves2012find}
Gon\c{c}alves, M., Cardoso, C., and Badke-Schaub, P. (2012).
\newblock Find your inspiration: exploring different levels of abstraction in
  textual stimuli.
\newblock In {\em DS 73-1 Proceedings of the 2nd International Conference on
  Design Creativity Volume 1}.

\bibitem[Gosline et~al., 1999]{gosline1999mechanical}
Gosline, J., Guerette, P., Ortlepp, C., and Savage, K. (1999).
\newblock The mechanical design of spider silks: from fibroin sequence to
  mechanical function.
\newblock {\em Journal of Experimental Biology}, 202(23):3295--3303.

\bibitem[Green et~al., 2014]{green2014crowd}
Green, M., Seepersad, C.~C., and H{\"o}ltt{\"a}-Otto, K. (2014).
\newblock Crowd-sourcing the evaluation of creativity in conceptual design: A
  pilot study.
\newblock In {\em ASME 2014 International Design Engineering Technical
  Conferences and Computers and Information in Engineering Conference}, pages
  V007T07A016--V007T07A016. American Society of Mechanical Engineers.

\bibitem[Griffiths and Steyvers, 2004]{griffiths2004finding}
Griffiths, T.~L. and Steyvers, M. (2004).
\newblock Finding scientific topics.
\newblock {\em Proceedings of the National Academy of Sciences}, 101(suppl
  1):5228--5235.

\bibitem[Hey et~al., 2008]{hey2008analogies}
Hey, J., Linsey, J., Agogino, A.~M., and Wood, K.~L. (2008).
\newblock Analogies and metaphors in creative design.
\newblock {\em International Journal of Engineering Education}, 24(2):283.

\bibitem[Hofmann, 1999]{hofmann1999probabilistic}
Hofmann, T. (1999).
\newblock Probabilistic latent semantic indexing.
\newblock In {\em Proceedings of the 22nd Annual International ACM SIGIR
  Conference on Research and Development in Information Retrieval}, pages
  50--57. ACM.

\bibitem[Jansson and Smith, 1991]{jansson1991design}
Jansson, D.~G. and Smith, S.~M. (1991).
\newblock Design fixation.
\newblock {\em Design studies}, 12(1):3--11.

\bibitem[Jur{\v{s}}i{\v{c}} et~al., 2012a]{jurvsivc2012bisociative}
Jur{\v{s}}i{\v{c}}, M., Cestnik, B., Urban{\v{c}}i{\v{c}}, T., and
  Lavra{\v{c}}, N. (2012a).
\newblock Bisociative literature mining by ensemble heuristics.
\newblock In {\em Bisociative Knowledge Discovery}, pages 338--358. Springer.

\bibitem[Jur{\v{s}}i{\v{c}} et~al., 2012b]{jurvsivc2012bridging}
Jur{\v{s}}i{\v{c}}, M., Sluban, B., Cestnik, B., Gr{\v{c}}ar, M., and
  Lavra{\v{c}}, N. (2012b).
\newblock Bridging concept identification for constructing information networks
  from text documents.
\newblock In {\em Bisociative Knowledge Discovery}, pages 66--90. Springer.

\bibitem[Kamahara et~al., 2005]{kamahara2005community}
Kamahara, J., Asakawa, T., Shimojo, S., and Miyahara, H. (2005).
\newblock A community-based recommendation system to reveal unexpected
  interests.
\newblock In {\em Multimedia Modelling Conference, 2005. MMM 2005. Proceedings
  of the 11th International}, pages 433--438. IEEE.

\bibitem[Kang, 2016]{kang2016bisociative}
Kang, S.~W. (2016).
\newblock Bisociative design: concept exploration by mining design associations
  across multiple products.

\bibitem[Kang and Tucker, 2017]{kang2017exploring}
Kang, S.~W. and Tucker, C.~S. (2017).
\newblock Exploring the correlation between new function attributes mined from
  different product domains and market sales.
\newblock {\em The Engineering Economist}, pages 1--30.

\bibitem[Kittur, 2010]{kittur2010crowdsourcing}
Kittur, A. (2010).
\newblock Crowdsourcing, collaboration and creativity.
\newblock {\em ACM Crossroads}, 17(2):22--26.

\bibitem[Kittur et~al., 2013]{Kittur:2013:FCW:2441776.2441923}
Kittur, A., Nickerson, J.~V., Bernstein, M., Gerber, E., Shaw, A., Zimmerman,
  J., Lease, M., and Horton, J. (2013).
\newblock The future of crowd work.
\newblock In {\em Proceedings of the 2013 Conference on Computer Supported
  Cooperative Work}, CSCW '13, pages 1301--1318, New York, NY, USA. ACM.

\bibitem[Koestler, 1964]{koestler1964act}
Koestler, A. (1964).
\newblock The act of creation.

\bibitem[Krestel et~al., 2009]{krestel2009latent}
Krestel, R., Fankhauser, P., and Nejdl, W. (2009).
\newblock Latent dirichlet allocation for tag recommendation.
\newblock In {\em Proceedings of the third ACM conference on Recommender
  systems}, pages 61--68. ACM.

\bibitem[Maaten and Hinton, 2008]{maaten2008visualizing}
Maaten, L. v.~d. and Hinton, G. (2008).
\newblock Visualizing data using t-sne.
\newblock {\em Journal of Machine Learning Research}, 9(Nov):2579--2605.

\bibitem[MATLAB, 2016]{MATLAB:2016}
MATLAB (2016).
\newblock {\em version 9.0.0 (R2016a)}.
\newblock The MathWorks Inc., Natick, Massachusetts.

\bibitem[Newman et~al., 2010]{newman2010visualizing}
Newman, D., Baldwin, T., Cavedon, L., Huang, E., Karimi, S., Martinez, D.,
  Scholer, F., and Zobel, J. (2010).
\newblock Visualizing search results and document collections using topic maps.
\newblock {\em Web Semantics: Science, Services and Agents on the World Wide
  Web}, 8(2):169--175.

\bibitem[Newman et~al., 2011]{newman2011improving}
Newman, D., Bonilla, E.~V., and Buntine, W. (2011).
\newblock Improving topic coherence with regularized topic models.
\newblock In {\em Advances in neural information processing systems}, pages
  496--504.

\bibitem[Onuma et~al., 2009]{onuma2009tangent}
Onuma, K., Tong, H., and Faloutsos, C. (2009).
\newblock Tangent: a novel,'surprise me', recommendation algorithm.
\newblock In {\em Proceedings of the 15th ACM SIGKDD international conference
  on Knowledge discovery and data mining}, pages 657--666. ACM.

\bibitem[Pahl et~al., 2007]{pahl2007generic}
Pahl, A.-K., Newnes, L., and McMahon, C. (2007).
\newblock A generic model for creativity and innovation: overview for early
  phases of engineering design.
\newblock {\em Journal of Design Research}, 6(1-2):5--44.

\bibitem[Pang and Seepersad, 2016]{pang2016crowdsourcing}
Pang, M.~A. and Seepersad, C.~C. (2016).
\newblock Crowdsourcing the evaluation of design concepts with empathic
  priming.
\newblock In {\em ASME 2016 International Design Engineering Technical
  Conferences and Computers and Information in Engineering Conference}, pages
  V007T06A004--V007T06A004. American Society of Mechanical Engineers.

\bibitem[Petri{\u{e}} et~al., 2009]{petriue2009literature}
Petri{\u{e}}, I., Urban{\u{e}}i{\u{e}}, T., Cestnik, B., and
  Macedoni-Luk{\v{s}}i{\u{e}}, M. (2009).
\newblock Literature mining method rajolink for uncovering relations between
  biomedical concepts.
\newblock {\em Journal of Biomedical Informatics}, 42(2):219--227.

\bibitem[Poincar{\'e}, 1910]{poincare1910mathematical}
Poincar{\'e}, H. (1910).
\newblock Mathematical creation.
\newblock {\em The Monist}, pages 321--335.

\bibitem[Purcell and Gero, 1996]{purcell1996design}
Purcell, A.~T. and Gero, J.~S. (1996).
\newblock Design and other types of fixation.
\newblock {\em Design studies}, 17(4):363--383.

\bibitem[Roberts, 1989]{roberts1989serendipity}
Roberts, R.~M. (1989).
\newblock Serendipity: Accidental discoveries in science.
\newblock {\em Serendipity: Accidental Discoveries in Science, by Royston M.
  Roberts, pp. 288. ISBN 0-471-60203-5. Wiley-VCH, June 1989.}, page 288.

\bibitem[Rosen-Zvi et~al., 2004]{rosen2004author}
Rosen-Zvi, M., Griffiths, T., Steyvers, M., and Smyth, P. (2004).
\newblock The author-topic model for authors and documents.
\newblock In {\em Proceedings of the 20th conference on Uncertainty in
  Artificial Intelligence}, pages 487--494. AUAI Press.

\bibitem[Salton and Buckley, 1988]{salton1988term}
Salton, G. and Buckley, C. (1988).
\newblock Term-weighting approaches in automatic text retrieval.
\newblock {\em Information Processing \& Management}, 24(5):513--523.

\bibitem[Schmidt et~al., 2012]{schmidt2012bisociative}
Schmidt, O., Kranjc, J., Mozeti{\v{c}}, I., Thompson, P., and Dubitzky, W.
  (2012).
\newblock Bisociative exploration of biological and financial literature using
  clustering.
\newblock In {\em Bisociative Knowledge Discovery}, pages 438--451. Springer.

\bibitem[Segond and Borgelt, 2009]{segond2009bisonet}
Segond, M. and Borgelt, C. (2009).
\newblock “bisonet” generation using textual data.
\newblock In {\em Workshop on Explorative Analytics of Information Networks at
  ECML PKDD 2009}, page~12.

\bibitem[Shah et~al., 2003]{Shah2003111}
Shah, J.~J., Smith, S.~M., and Vargas-Hernandez, N. (2003).
\newblock Metrics for measuring ideation effectiveness.
\newblock {\em Design Studies}, 24(2):111 -- 134.

\bibitem[Siangliulue et~al., 2015]{siangliulue2015toward}
Siangliulue, P., Arnold, K.~C., Gajos, K.~Z., and Dow, S.~P. (2015).
\newblock Toward collaborative ideation at scale: Leveraging ideas from others
  to generate more creative and diverse ideas.
\newblock In {\em Proceedings of the 18th ACM Conference on Computer Supported
  Cooperative Work \& Social Computing}, pages 937--945. ACM.

\bibitem[Stober et~al., 2012]{stober2012bisociative}
Stober, S., Haun, S., and N{\"u}rnberger, A. (2012).
\newblock Bisociative music discovery and recommendation.
\newblock In {\em Bisociative Knowledge Discovery}, pages 472--483. Springer.

\bibitem[Swanson, 1986]{swanson1986fish}
Swanson, D.~R. (1986).
\newblock Fish oil, raynaud's syndrome, and undiscovered public knowledge.
\newblock {\em Perspectives in biology and medicine}, 30(1):7--18.

\bibitem[Teh et~al., 2004]{teh2004sharing}
Teh, Y.~W., Jordan, M.~I., Beal, M.~J., and Blei, D.~M. (2004).
\newblock Sharing clusters among related groups: Hierarchical dirichlet
  processes.
\newblock In {\em NIPS}, pages 1385--1392.

\bibitem[Tucker and Kang, 2012]{tucker2012bisociative}
Tucker, C. and Kang, S. (2012).
\newblock Bisociative design framework for knowledge discovery across seemingly
  unrelated product domains.
\newblock {\em Proceedings of the ASME IDETC/CIE, Chicago, IL, Paper No.
  DETC2012-70764}.

\bibitem[Wei and Croft, 2006]{wei2006lda}
Wei, X. and Croft, W.~B. (2006).
\newblock {LDA}-based document models for ad-hoc retrieval.
\newblock In {\em Proceedings of the 29th Annual International ACM SIGIR
  conference on Research and development in information retrieval}, pages
  178--185. ACM.

\bibitem[Zahner et~al., 2010]{zahner2010fix}
Zahner, D., Nickerson, J.~V., Tversky, B., Corter, J.~E., and Ma, J. (2010).
\newblock A fix for fixation? rerepresenting and abstracting as creative
  processes in the design of information systems.
\newblock {\em AI EDAM}, 24(2):231--244.

\end{thebibliography}

\end{document}